 \documentclass[journal, comsoc]{IEEEtran}

\usepackage{gensymb}
\usepackage{cite}
\usepackage{amsmath,amssymb,amsfonts}
\usepackage{graphicx}
\usepackage{xcolor}
\usepackage{kotex}  
\usepackage{caption}
\usepackage{subcaption}
\usepackage{caption}
\usepackage{subcaption}
\usepackage{tabularx}
\usepackage{booktabs}

\usepackage{stfloats}
\usepackage{float}
\usepackage{wrapfig}
\usepackage{xcolor}
\usepackage{colortbl} 
\usepackage{color,soul}
\usepackage{url}
\usepackage{verbatim}
\usepackage{graphicx}
\usepackage{cite}
\usepackage{adjustbox}

\usepackage[acronym,toc,shortcuts]{glossaries}
\newacronym{BS}{BS}{Base Stations}
\newacronym{NTN}{NTN}{Non-Terrestrial Networks}
\newacronym{TN}{TN}{Terrestrial Networks}

\newacronym{NOMA}{NOMA}{Non-Orthogonal Multiple Access}
\newacronym{IoT}{IoT}{Internet of Things}
\newacronym{UEs}{UEs}{User Equipment's}
\newacronym{UE}{UE}{User Equipment}
\newacronym{UAV}{UAV}{Uncrewed Aerial Vehicles}
\newacronym{SDG}{SDG}{Sustainable Development Goals}
\newacronym{UN}{UN}{United Nations}
\newacronym{HAPS}{HAPS}{High-Altitude Platform Stations}
\newacronym{AI}{AI}{Artificial Intelligence}
\newacronym{EE}{EE}{Energy Efficiency}
\newacronym{3GPP}{3GPP}{3$^{\text{rd}}$ Generation Partnership Project}
\newacronym{THz}{THz}{Terahertz}
\newacronym{mmWave}{mmWave}{millimeter Wave}
\newacronym{PLMN}{PLMN}{Public Land Mobile Networks}
\newacronym{MNO}{MNO}{Mobile Network Operators}
\newacronym{LEO}{LEO}{Low Earth Orbit}
\newacronym{KPI}{KPI}{Key Performance Indicators}
\newacronym{GEO}{GEO}{Geostationary Equatorial Orbit}
\newacronym{MEO}{MEO}{Medium-Earth Orbit}
\newacronym{CapEx}{CapEx}{Capital Expenditures}
\newacronym{IMS}{IMS}{IP Multimedia Subsystem}
\newacronym{SA}{SA}{System and Architecture}
\newacronym{RAN}{RAN}{Radio Access Network}
\newacronym{NGSO}{NGSO}{Non-Geostationary Satellite Orbit}
\newacronym{UPF}{UPF}{User Plane Function}
\newacronym{mMIMO}{mMIMO}{massive Multiple Input Multiple Output}
\newacronym{QoS}{QoS}{Quality-of-Service}
\newacronym{LAN}{LAN}{Local Area Network}
\newacronym{RHS}{RHS}{Reconfigurable Holographic Surfaces}
\newacronym{ISAC}{ISAC}{Integrated Sensing and Communication}
\newacronym{SNO}{SNO}{Satellite Network Operators}
\newacronym{MENA}{MENA}{Middle East and North Africa}
\newacronym{LTE}{LTE}{Long-Term Evolution}
\newacronym{NAS}{NAS}{Non Access Stratum}
\newacronym{Wi-Fi}{Wi-Fi}{Wireless Fidelity}
\newacronym{FFR}{FFR}{Full Frequency Reuse}
\newacronym{ZF}{ZF}{Zero Forcing}
\newacronym{MIMO}{MIMO}{Multiple-Input-Multiple-Output}
\newacronym{MMSE}{MMSE}{Minimum Mean Square Error}
\newacronym{SSB}{SSB}{Synchronisation Signal Blocks}
\newacronym{CSI}{CSI}{Channel State Information}
\newacronym{CQI}{CQI}{Channel Quality Indicator}
\newacronym{RS}{RS}{Reference Signal}
\newacronym{GNSS}{GNSS}{Global Navigation Satellite System}
\newacronym{AMF}{AMF}{Access and Mobility Management Function}
\newacronym{LMF}{LMF}{Location Management Function}
\newacronym{5GC}{5GC}{5G Core Network}
\newacronym{NFC}{NFC}{Near-Field Communication}
\newacronym{NC-JT}{NC-JT}{Non-Coherent JT}
\newacronym{MC}{MC}{Multi-Connectivity}
\newacronym{RRM}{RRM}{Radio Resource Management}
\newacronym{ML}{ML}{Machine Learning}
\newacronym{RU}{RU}{Radio Unit}
\newacronym{CU}{CU}{Control Unit}
\newacronym{DU}{DU}{Distributed Unit}
\newacronym{RIC}{RIC}{RAN Intelligent Controller}
\newacronym{CN}{CN}{Core Network}
\newacronym{EIRP}{EIRP}{Equivalent Isotropic Radiated Power}
\newacronym{VSAT}{VSAT}{Very Small Aperture Terminal}
\newacronym{RIS}{RIS}{Reconfigurable Intelligent Surfaces}
\newacronym{CR}{CR}{Cognitive Radios}
\newacronym{ELAA}{ELAA}{Extremely Large Aperture Arrays}
\newacronym{gNB}{gNB}{Next-Generation Node B}
\usepackage{multirow}
\usepackage{array,boldline, makecell,booktabs}
\usepackage{longtable} 

    \usepackage[group-separator={,},group-minimum-digits=4]{siunitx}

\begin{document}
%
\title{Reconfigurable Holographic Surfaces and Near Field Communication for Non-Terrestrial Networks: Potential and Challenges}
%
%
%

\author{Muhammad Ali Jamshed,~\IEEEmembership{Senior Member,~IEEE,}, Muhammad Ahmed Mohsin~\IEEEmembership{Graduate Member,~IEEE,}, Hongliang Zhang, Bushra Haq, Aryan Kaushik~\IEEEmembership{Member,~IEEE}, Boya Di, and Weiwei Jiang
\thanks{
M. A. Jamshed is with the College of Science and Engineering, University	of Glasgow, UK (e-mail: muhammadali.jamshed@glasgow.ac.uk).\\ 
M. A. Mohsin is with the Department of Electrical Engineering, Stanford University, Stanford, CA, USA (email: muahmed@stanford.edu).\\
H. Zhang is with the School of Electronics, Peking University, China (email: hongliang.zhang@pku.edu.cn)\\
B. Haq is with Balochistan University of Information Technology, Engineering and Management Sciences, Pakistan (e-mail: bushra.haq@buitms.edu.pk).\\
A. Kaushik is with the Department of Computing \& Mathematics, Manchester Metropolitan University, UK (e-mail: a.kaushik@mmu.ac.uk).\\
B. Di is with the School of Electronics, Peking University, China (email: boya.di@pku.edu.cn)
\\
W. Jiang is with the Beijing University of Posts and Telecommunications, China (email: jww@bupt.edu.cn)
\\
}
}

%
%

\markboth{Submitted to IEEE Wireless Communication Magazine}%
{Shell \MakeLowercase{\textit{et al.}}: Bare Demo of IEEEtran.cls for IEEE Journals}

\maketitle

\begin{abstract}
To overcome the challenges of ultra-low latency, ubiquitous coverage, and soaring data rates, this article presents a combined use of Near Field Communication (NFC) and Reconfigurable Holographic Surfaces (RHS) for Non-Terrestrial Networks (NTN). A system architecture has been presented, which shows that the integration of RHS with NTN platforms such as satellites, High Altitute Platform Stations (HAPS), and Uncrewed Aerial Vehicles (UAV) can achieve precise beamforming and intelligent wavefront control in near-field regions, enhancing Energy Efficiency (EE), spectral utilization, and spatial resolution. Moreover, key applications, challenges, and future directions have been identified to fully adopt this integration. In addition, a use case analysis has been presented to improve the EE of the system in a public safety use case scenario, further strengthening the UAV-RHS fusion. 


\end{abstract}


\IEEEpeerreviewmaketitle


\section{Introduction} 
As 5G wireless networks continue to be deployed globally, the focus of the research community is increasingly shifting toward the conceptualization and realization of 6G communication systems. Although 5G has introduced significant advancements such as \ac{mmWave} technology and \ac{mMIMO}, enabling applications in smart cities and the \ac{IoT}, it still does not meet the demands posed by emerging technologies such as virtual reality, autonomous vehicles, and the metaverse. Consequently, 6G aims to overcome these limitations by providing significantly higher data rates, ultra-low latency, improved reliability, and ubiquitous coverage across terrestrial, aerial, and space-based networks, all while incorporating intelligent adaptive functionalities \cite{mu2024reconfigurable}.

To achieve these ambitious objectives, several advanced technologies are being explored, including extremely large-scale MIMO (XL-MIMO), \ac{THz} communications, \ac{RIS}, and \ac{NTN}. XL-MIMO and \ac{THz} communications promise unprecedented data throughput, but introduce new complexities related to high-frequency propagation and expanded antenna arrays. These, in turn, affect the Rayleigh distance, pushing more users into the near-field region of \ac{BS}, where traditional far-field propagation models no longer apply due to the spherical nature of wavefronts. This fundamental shift necessitates a re-evaluation of existing communication theories and models to accommodate near-field effects in 6G systems.

Among the emerging technologies, \ac{RIS} has gained particular attention for its potential to shape and optimize the wireless propagation environment \cite{jamshed2024synergizing}. Comprising a dense array of low-cost, nearly passive electromagnetic elements, \ac{RIS} can enhance signal strength and suppress interference through intelligent wave manipulation. Parallel to \ac{RIS}, \ac{RHS} have emerged as a promising solution to support ultra-mMIMO deployments, a key requirement for 6G. \ac{RHS} utilizes metamaterial-based antennas with simple components such as diodes, enabling the control of electromagnetic wavefronts with significantly lower power consumption and reduced hardware complexity. This makes \ac{RHS} an ideal candidate to implement large-scale antenna arrays necessary for high-frequency, high-capacity wireless communication, while also addressing the energy and cost constraints associated with conventional phased array systems \cite{deng2021reconfigurable}.

Achieving the vision of a fully integrated, intelligent and ubiquitous 6G network also requires the incorporation of \ac{NTN}. \ac{NTN} include satellite systems and aerial platforms such as \ac{UAV} and \ac{HAPS} \cite{jamshed2025tutorial, 10847914}. These platforms serve as relay nodes or \ac{BS} that extend connectivity to underserved or remote areas, thus playing a pivotal role in ensuring uninterrupted global communication coverage. However, integrating \ac{TN} and \ac{NTN} introduces several technical challenges, such as Doppler shift, latency, and \ac{EE}. In this context, the synergy between \ac{RHS} and \ac{NTN}, particularly through the deployment of \ac{RHS} on aerial platforms, presents a compelling approach to improve performance while maintaining sustainability. The combination enables intelligent beamforming, efficient resource allocation, and adaptive network configurations, thereby supporting the goals of green communications and carbon neutrality.
To advance this field, it is essential to address a number of critical questions regarding the use of \ac{RHS} and \ac{NFC} for \ac{NTN} in the context of 6G systems. Q1: What practical strategies exist for implementing \ac{RHS}-enhanced \ac{NFC} through \ac{NTN} platforms? Q2: In what ways does the utilization of \ac{RHS} and \ac{NFC} for \ac{NTN} contribute to \ac{EE} and support carbon neutrality goals? Q3: How can the combined use of \ac{RHS} and \ac{NFC} for \ac{NTN} overcome key challenges in 6G, including scalability, reliability and latency? Q4: What new applications and use cases can be enabled through this integration of emerging technologies?

This paper seeks to explore and answer these questions by providing a comprehensive analysis of the current landscape, identifying technological gaps, and proposing frameworks that leverage the strengths of \ac{NFC}, \ac{RHS}, and \ac{NTN} to enable the next generation of wireless connectivity. Moreover, the contribution to the knowledge of this article is summarized as follows.

\begin{itemize}
    \item We provide an extensive overview of \ac{NFC}, \ac{RHS}, and \ac{NTN} technologies.  
    \item We provide a system architecture design for \ac{RHS}-enhanced \ac{NFC} through \ac{NTN} platforms.
    \item We provide a thorough analysis of the applications and challenges associated with this trifecta.
    \item We provide a use case demonstration of how \ac{EE} of wireless network could be improved by integrating \ac{RHS} with aerial \ac{NTN}.
    \item We provide a future roadmap of technologies to succeed in adopting the combined use of \ac{NFC}, \ac{RHS} and \ac{NTN} for future 6G systems.
\end{itemize}
The rest of the article is organized as follows. Section II provides an extensive overview of \ac{NFC}, \ac{RHS}, and \ac{NTN} technologies. Section III provides applications of  \ac{RHS} and \ac{NFC} for \ac{NTN} and challenges associated with them. Section IV demonstrates a use case analysis of improving the \ac{EE} of the \ac{RHS} assisted aerial \ac{NTN} for a public safety scenario. Section V provides an overview of future research challenges; finally, we conclude the paper in Section VII.



\section{Revisiting: NFC, RHS, and Aerial NTN}

Recently, \ac{3GPP} Release 20 introduced key advancements for 5G-Advanced and also marked the first official step toward defining study items for 6G. Building on the foundation laid in Release 20, integrated, intelligent, and ubiquitous connectivity remains a cornerstone of the 6G development \cite{lin2025tale,jamshed2025non}. To realize this vision, technologies such as \ac{RHS}, \ac{NFC}, and \ac{NTN} play a critical role. In this section, we provide a high-level overview of these three technologies and explore how their integration can help bring 6G to fruition. 

\subsection{What is Reconfigurable Holographic surface}

The \ac{RHS} functions as a specialized leaky-wave antenna. An incident electromagnetic wave travels along a guided structure within the \ac{RHS}, where engineered surface discontinuities modulate it into a leaky-wave that radiates into free space. Coordinated radiation from these points forms directional beams. The \ac{RHS} consists of three key components, as illustrated in Fig. \ref{fig:1}: a feed system, a waveguide, and metamaterial-based radiating elements. Embedded within the bottom layer, the feed system launches reference waves that excite the surface field, enabling a thinner profile compared to traditional arrays with bulky external feeds. The waveguide efficiently channels these waves across the surface, while the radiating elements, engineered from tunable metamaterials, control the beam direction through electric or magnetic biasing. The interaction between the reference wave and these elements dictates the overall radiation pattern \cite{deng2021reconfigurable}.

Alongside \ac{RHS}, \ac{RIS} has emerged as another cutting-edge solution in metasurface-based wireless communication. Composed of tunable metamaterial units, \ac{RIS} reflects incident signals and forms directed beams toward the target users. \ac{RIS} technology plays a pivotal role in shaping intelligent radio environments. Recent research has explored hybrid beamforming with \ac{RIS} in multi-user systems. While both \ac{RHS} and \ac{RIS} enable beamforming, they differ significantly in structure and operation, as outlined in Table \ref{table1}.
\begin{table*}[t!]
\centering
\begin{tabularx}{\linewidth}{|l|X|X|}
\hline
\textbf{Parameters} & \textbf{RIS} & \textbf{RHS} \\
\hline
Operation Type & Passive or Semi-passive & Active \\
\hline
Beamforming Precision & Moderate & High \\
\hline
Adaptive Beam Shaping & Limited & High \\
\hline
Channel Adaptability & Lower Flexibility & Higher Flexibility \\
\hline
Interference Suppression & Moderate & High \\
\hline
Spectral Efficiency & Moderate & High \\
\hline
Energy Efficiency & Low & High \\
\hline
Wave Manipulation & Phase Control & Phase \& Amplitude Control \\
\hline
MIMO Support & Indirect & Direct \\
\hline
Deployment Cost & Low & High \\
\hline
Hardware Complexity & Low & High \\
\hline
Signal Processing Demand & Low & High \\
\hline
Coverage & Limited & High \\
\hline
\end{tabularx}
\caption{Comparative analysis of RIS versus RHS}
\label{table1}
\end{table*}

\subsection{What is Near Field Communication}

The adoption of emerging 6G technologies, such as \ac{THz} communication and XL-MIMO, has significantly increased carrier frequencies and antenna array apertures. These developments extend the Rayleigh (Fraunhofer) distance, which defines the boundary between near-field and far-field electromagnetic propagation. For example, a uniform linear array with a 0.5 meter aperture operating at 28 GHz has a Rayleigh distance of approximately 47 meters, expanding the near-field region and increasing the likelihood that user terminals fall within it. This shift invalidates conventional far-field assumptions, as the wavefront curvature becomes non-negligible, making plane wave approximations unreliable.

\begin{figure}[t!]
    \centering
    \includegraphics[width=\linewidth]{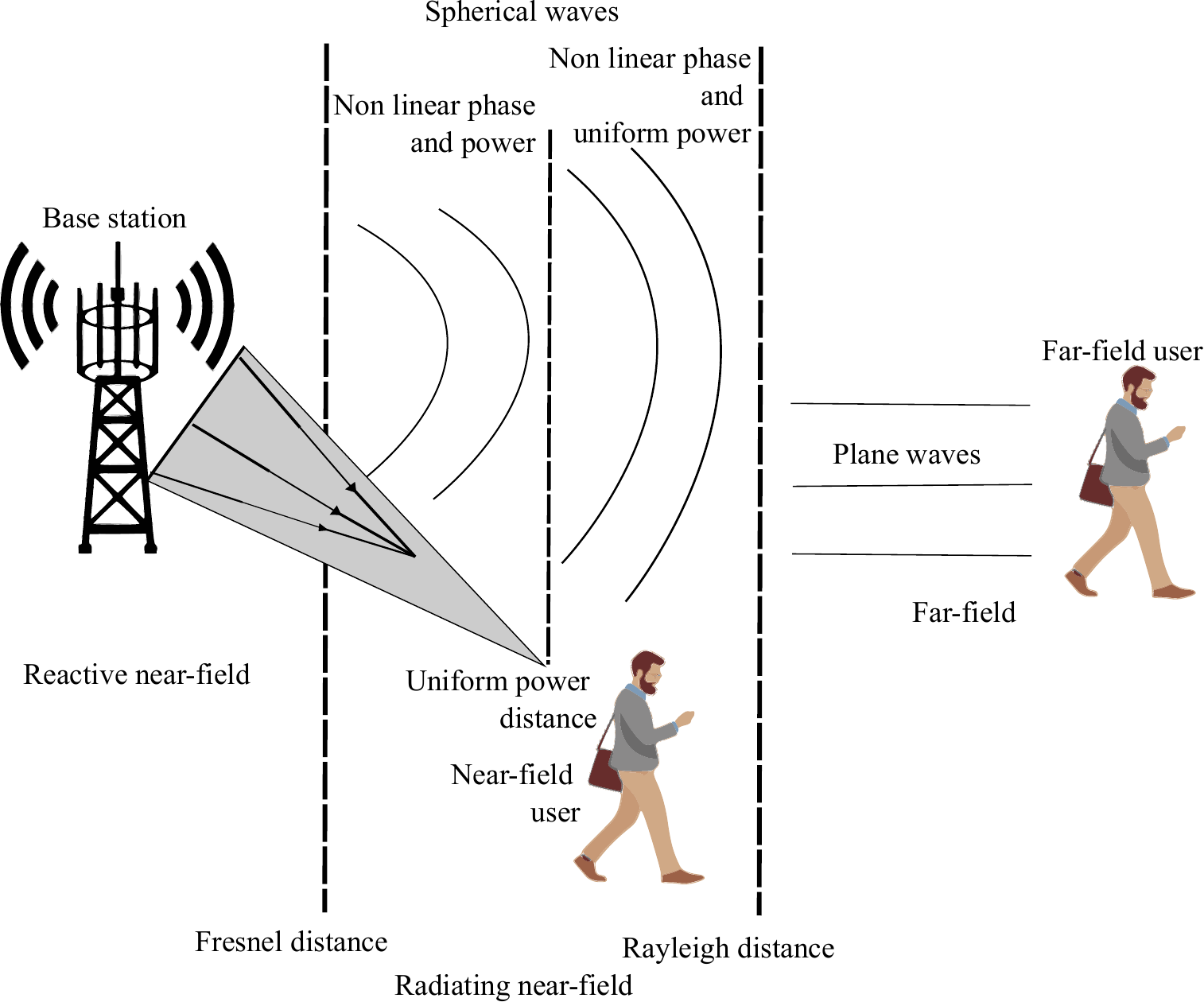}
    \caption{An illustration of NFC}
    \label{fig:2}
\end{figure}

In the near-field, electromagnetic propagation must be modeled more precisely. As shown in Fig. \ref{fig:2} (illustrating an uplink scenario with a \ac{BS}, though the analysis also applies to downlink cases), the conventional Uniform Planar Wave (UPW) model used in far-field conditions assumes linear phase shifts and uniform power across antenna elements. However, in the near-field, the Uniform Spherical Wave (USW) model applies when the distance is below the Rayleigh limit but above the non-uniform power threshold; here, the phase becomes non-linear while power remains uniformly distributed. When the distance drops further, the Non-uniform Spherical Wave (NUSW) model becomes necessary, accounting for both non-linear phase variations and non-uniform power across the array \cite{mu2024reconfigurable}. In the far-field, wavefronts are effectively planar, allowing phase to be modeled linearly with respect to the antenna index, facilitating traditional beamsteering based on Angle of Arrival (AoA). In contrast, near-field conditions require a non-linear phase model dependent on both angle and distance, enabling precise beam-focusing at specific spatial points. This paradigm shift necessitates rethinking channel models and system design strategies for 6G, highlighting the importance of accurately modeling near-field behavior to fully exploit its potential for spatial resolution and energy focus.

\subsection{What are Non-Terrestrial Networks}
According to \ac{3GPP} Release 17, \ac{NTN} primarily involve satellite systems, with \ac{UAV} and \ac{HAPS} recognized as specialized use cases. While satellites remain central to \ac{NTN}, offering persistent coverage and broad communication capabilities, the concept also encompasses airborne platforms that deliver connectivity on demand. These technologies play a crucial role in extending communication services to remote, underserved, or disaster-affected regions where conventional \ac{TN} are impractical. The global \ac{NTN} market is projected to reach 30 billion dollars by 2030, with significant momentum driven by the government, defense, and public safety sectors, accounting for more than 35\% of the growth. As the demand for ubiquitous and resilient connectivity grows, \ac{NTN} are transitioning from niche applications to mainstream deployment. Their utility spans critical scenarios such as emergency response, disaster relief, search and rescue operations, and public safety communications.

In the context of emerging 6G networks, \ac{NTN} are becoming pivotal components. \ac{NTN} functioning as aerial/space \ac{BS}, mobile relays, or even user equipment, can extend coverage to areas where infrastructure is limited or temporarily unavailable, as illustrated in Fig.\ref{fig11}. The use of \ac{UAV} (a special use case of \ac{NTN}) allows for dynamic repositioning, enhancing network adaptability and efficiency, particularly in emergencies or large-scale public events. \ac{NTN} contribute significantly to 6G performance targets, such as ultra-low latency and high reliability, essential for applications like autonomous systems, real-time remote surgery, and industrial automation. Unlike traditional 2D networks, cooperative \ac{NTN} enable 3D architectures, improving spectrum utilization and resource management by using vertical space.


\subsection{RHS and NFC for NTN}

Based on the insights from the preceding subsections, \ac{NFC}, \ac{RHS}, and \ac{NTN}, each offer distinct yet complementary advantages critical for realizing the full vision of 6G. As shown in Fig.~\ref{fig:2}, these technologies can be synergistically integrated to enable intelligent, ubiquitous, and adaptive wireless connectivity across diverse environments. For example, \ac{RHS}-enabled terrestrial infrastructure can leverage \ac{NFC} to achieve ultra-precise beamfocusing, facilitating high-capacity, low-latency links to low-altitude \ac{UAV} functioning within the near-field region. These \ac{UAV}, in turn, act as agile access points or relays, providing last-hop connectivity and localized edge computing capabilities in hard-to-reach or dynamic environments. Moreover, the Direct to Satellite (D2S), \ac{RHS}, and \ac{NFC} can work cooperatively to facilitate a user from multi-band utilization and provide seamless connectivity. Such integration supports a variety of use cases, from real-time positioning and tracking in autonomous delivery networks to delay-sensitive public safety communications during emergency response operations.

\ac{UAV} equipped with compact \ac{RHS} modules can dynamically form directional beams to serve both ground users and other aerial/space nodes with improved \ac{EE} and spectral reuse. When operating in disaster-affected or infrastructure-scarce regions, these aerial/space platforms can establish self-organizing \ac{NTN}, by cooperatively relying on \ac{HAPS} and satellites, ensuring communication continuity and network resilience. \ac{NFC} further enables fine-grained spatial awareness and beam adaptation, essential for managing near-field/far-field user heterogeneity in densely populated urban scenarios. Coupled with the programmable nature of \ac{RHS}, the system can intelligently adapt its transmission strategy in real time based on the location of the user, the mobility of the device and the service requirements. Overall, the integration of \ac{RHS}, \ac{NFC}, and \ac{NTN} forms a foundational triad for 6G networks, empowering intelligent surface-assisted communication, enhancing spectrum utilization, and extending seamless coverage across terrestrial and non-terrestrial domains.

\begin{figure*}[t!]
    \centering
    \includegraphics[width=\linewidth]{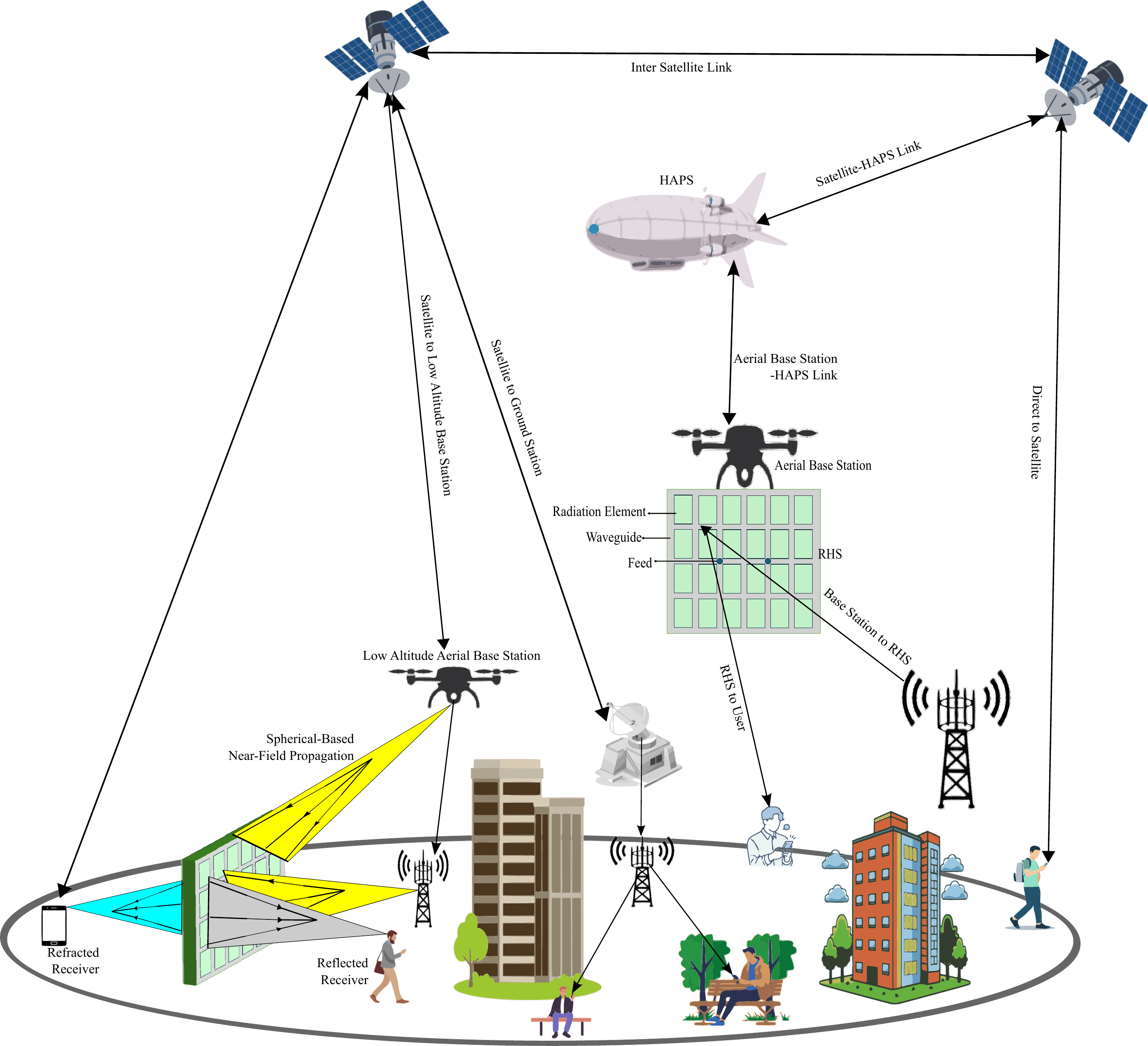}
    \caption{An illustration of RHS and NFC for integrated TN and NTN.}
    \label{fig:1}
\end{figure*}

\section{Applications and Challenges}
In this section, we discuss the applications and the challenges faced by the utilization of \ac{NFC} and \ac{RHS}, for \ac{NTN}. 
\subsection{Applications}
\subsubsection{Consumer Communication and Smart Environments}
Modern communication environments stand to benefit from \ac{NFC}, \ac{RHS}, and \ac{NTN} integration in daily consumer scenarios. For instance, at crowded events or smart cities, aerial platforms can serve as on-demand \ac{BS} to offload networks and extend coverage. \ac{RHS} embedded in buildings or vehicles intelligently beamform signals, focusing high-rate data streams toward user hotspots or Augmented Reality (AR) devices. Meanwhile, \ac{NFC} provides seamless local interactions, where a user can tap smartphone on \ac{NFC} point to quickly authenticate or retrieve context-specific data (e.g. digital tickets or AR content), which are then relayed via the drone-RHS network. 


\subsubsection{Industrial and IoT Applications} In industrial and logistics settings, combining \ac{NFC}, \ac{RHS}, and \ac{NTN} unlocks powerful capabilities. Autonomous drones equipped with \ac{NFC} readers can perform inventory tracking and asset monitoring in warehouses and factories. Field trials have shown drones autonomously scanning \ac{NFC} tags at up to 1,000 items per second with 99.9\% accuracy, far outperforming fixed readers. Moreover, \ac{RHS} can dynamically redirect or boost wireless signals around obstructions, creating smart propagation corridors for factory floor connectivity. In essence, \ac{NFC}-tagged equipment and products provide fine-grained data, \ac{UAV} serve as flexible readers and airborne routers, and \ac{RHS} units ensure the communication link is robust (by controlling reflections and nullifying interference). 



\subsubsection{Emergency Response and Public Safety} The integration of \ac{NFC}, \ac{RHS}, and \ac{NTN} is valuable in emergency and disaster scenarios. Post-disaster connectivity can be rapidly deployed using \ac{UAV}, whereas, a \ac{UAV} enhanced with \ac{RHS}, can form directed beams to reach survivors or emergency crews, and \ac{UAV} can rely the information using cooperative \ac{HAPS} and satellite networks, as shown in Fig. \ref{fig:1}. For example, an \ac{RHS}-equipped drone can steer a narrow beam to individuals buried under rubble who carry \ac{NFC}-enabled smartphones or ID tags. \ac{NFC} plays a critical role in identification and data exchange on the front lines: rescue workers can tap an accident victim’s \ac{NFC} medical ID to instantly retrieve health information, or drones can read \ac{NFC}-based tags on patients to prioritize aid. The data gathered via \ac{NFC} can be relayed through the cooperative \ac{NTN} network link to medical centers in real time. This combination thus facilitates ubiquitous emergency communication, combining short-range contactless data pickup with long-range aerial/space coverage and adaptive environment shaping for maximal reach and reliability.


\subsubsection{Ubiquitous Connectivity and Smart Infrastructure} Beyond specific scenarios, the interplay of \ac{NFC}, \ac{RHS}, and \ac{NTN} enables truly ubiquitous connectivity for the Internet of Everything (IoE). Imagine a smart city or large-scale environmental monitoring deployment: thousands of battery-free \ac{NFC} sensors (for example, wearable patches or micro-fliers) are distributed to monitor conditions like air quality, structural health, or crowd density. These devices operate in the near-field zone and periodically store data. A network of drones then acts as data mules, flying close to each sensor to harvest its data via \ac{NFC} links. \ac{RHS} further enrich this ubiquitous network by enabling spatially focused communication and power transfer. For instance, a programmable metasurface on a lamppost or on the drone itself could form a highly directed beam toward a cluster of passive \ac{NFC} sensors, either to energize them (wireless power) or to enhance the backscatter signal strength. Additionally, \ac{RHS} units in the infrastructure can dynamically re-route signals from aerial/space nodes to the fiber backhaul, effectively creating a seamless link from tiny local sensors to the core network. The synergy allows pervasive sensing and communication, hence, realizes an ambient intelligent network, supporting consumer comfort, smart infrastructure management, and environmental sustainability all at once.


\subsection{Challenges}
\subsubsection{Beam Coordination and Alignment} Coordinating beams across \ac{NFC}, \ac{RHS}, and \ac{NTN} components introduces significant challenges. Reconfigurable surfaces and aerial/space platforms must precisely align their beams to achieve the promised gains in coverage and capacity. This is non-trivial when the transmitter, reflector, and receiver might all be moving (e.g., a drone with \ac{RHS} serving a user device), and requires advanced sensing and control. Near-field beamforming adds another layer of complexity: unlike far-field beams, a focused near-field beam requires adjusting to the spherical wavefront and distance of the target. If \ac{UAV} moves from far-field into near-field of a device, the beam shape must transition accordingly. Although near-field beam focusing can mitigate interference and can support multiple high-data streams to \ac{UAV}, \ac{UAV} mobility makes continuous alignment difficult. In practice, this demands beam coordination algorithms that account for drone flight dynamics, fast channel variation, and the tuning delays of \ac{RHS} elements. Failing to coordinate beams could result in coverage holes or sudden drops in link quality. Thus, developing fast beam tracking and predictive alignment (potentially Artificial Intelligence (AI)-driven) is critical to unlock the full synergy of these technologies.

\subsubsection{Spectrum and Interference Management} 
The combination of \ac{NFC}, \ac{RHS} and \ac{NTN} spans a broad spectrum, from higher frequency bands to mmWave or \ac{THz} for high-speed aerial/space links. Managing this spectrum usage and avoiding interference is a considerable challenge. Spectrum management must ensure that the introduction of \ac{RHS} and aerial/space relays does not unduly pollute the airwaves \cite{mohsin2025hierarchicaldeepreinforcementlearning}. Traditional interference avoidance techniques must be enhanced with the new degrees of freedom: surfaces can be programmed to introduce nulls or attenuate in certain directions, and dynamic spectrum access could allocate distinct bands for \ac{NFC} interactions versus backhaul. Interference management algorithms that leverage AI to dynamically optimize spectrum use and power levels need to be explored. The near-field zone adds another twist: when using large intelligent surfaces or extremely high-frequency links, the distinction between near- and far-field propagation blurs, potentially causing interference regions that are highly localized.


\subsubsection{Security and Control} The integration of \ac{RHS} and drones into the communication infrastructure opens new security challenges. \ac{RHS} needs programming of its reflective elements, and drones obey wireless control and networking instructions. This introduces potential vulnerabilities: an attacker who gains control of the \ac{RHS} could redirect or siphon signals (violating user privacy or causing service disruption), while a hijacked \ac{UAV} could physically or digitally disrupt communications. Ensuring secure control channels and authentication for these devices is therefore critical. Thus, robust encryption and authentication are needed for the protocols that establish the metasurface states. Physical layer security techniques can be harnessed too, interestingly, near-field beamforming inherently adds security by confining signals to a smaller spatial region, which can reduce far-away eavesdropping risk.

\subsubsection{EE Constraints} In an integrated \ac{NFC}, \ac{RHS} and \ac{NTN} environment, energy is a limiting factor. \ac{UAV} are typically battery-powered and face strict flight time limits, particularly when carrying communication payloads or actively transmitting. Introducing \ac{RHS} hardware on a drone can further tax its energy reserves. Prolonged operation of \ac{UAV} as airborne \ac{BS} or readers, thus demands innovative power management, such as, use of energy harvesting, and optimizing \ac{UAV} trajectory for minimal power use. Reconfigurable surfaces themselves, if passive, consume little energy for beamforming; however, in \ac{RHS}, the control circuitry and active elements increase power usage. Moreover, \ac{NFC} presents its own energy considerations. Passive \ac{NFC} tags are power efficient but possess limited range. To extend \ac{NFC} functions via drones or larger areas, we might need higher power readers or specially designed metasurfaces to channel energy to tags, which could conflict with energy-saving goals. Without careful design, the combined system might offer great performance but at the cost of unsustainable energy use.





\begin{figure}[t!]
    \centering
    \includegraphics[width=\linewidth]{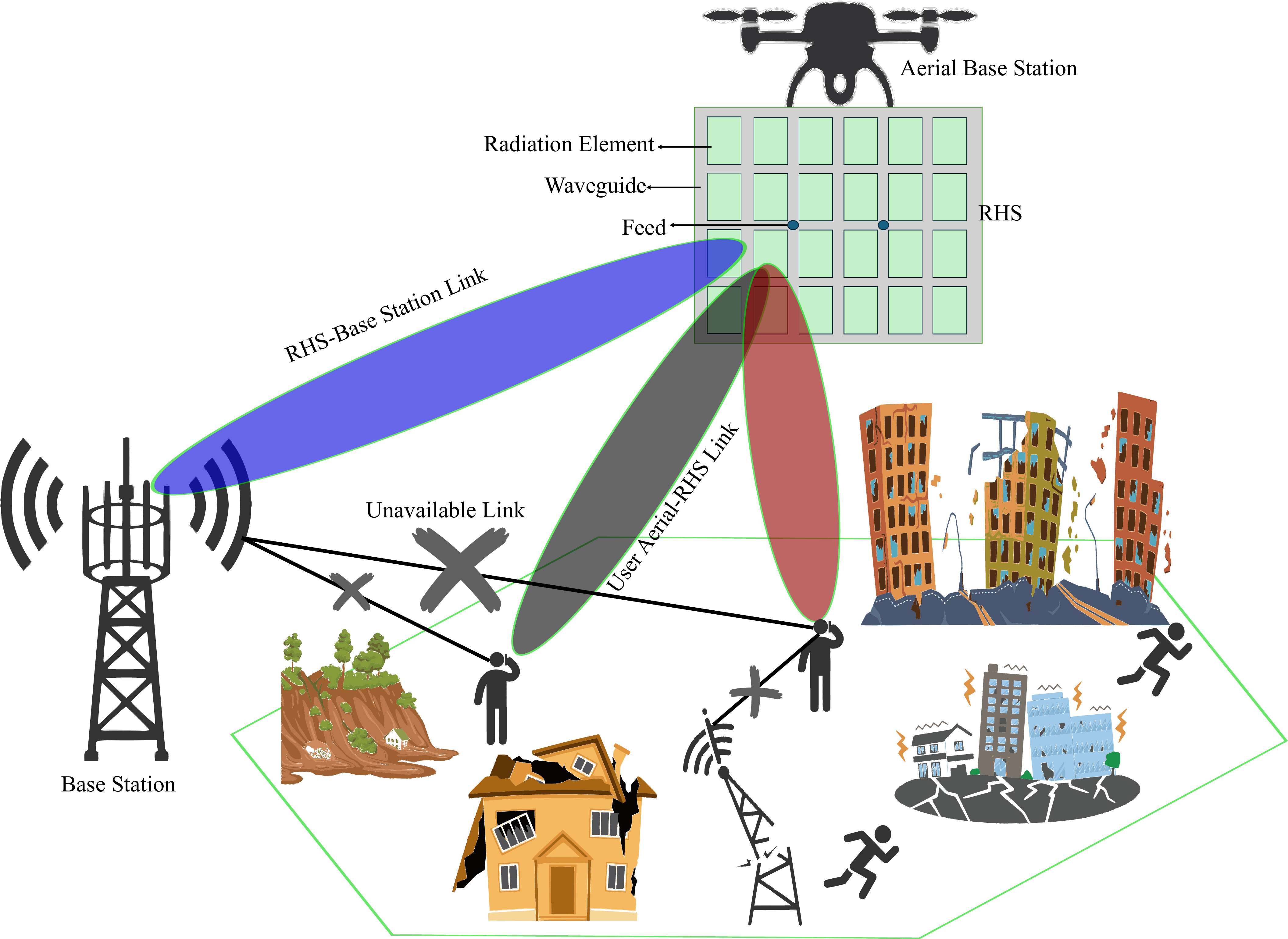}
    \caption{System model of the proposed fusion of UAV and RHS for public safety application.}
    \label{fig11}
\end{figure}

\begin{figure}[t!]
    \centering
    \includegraphics[width=\linewidth]{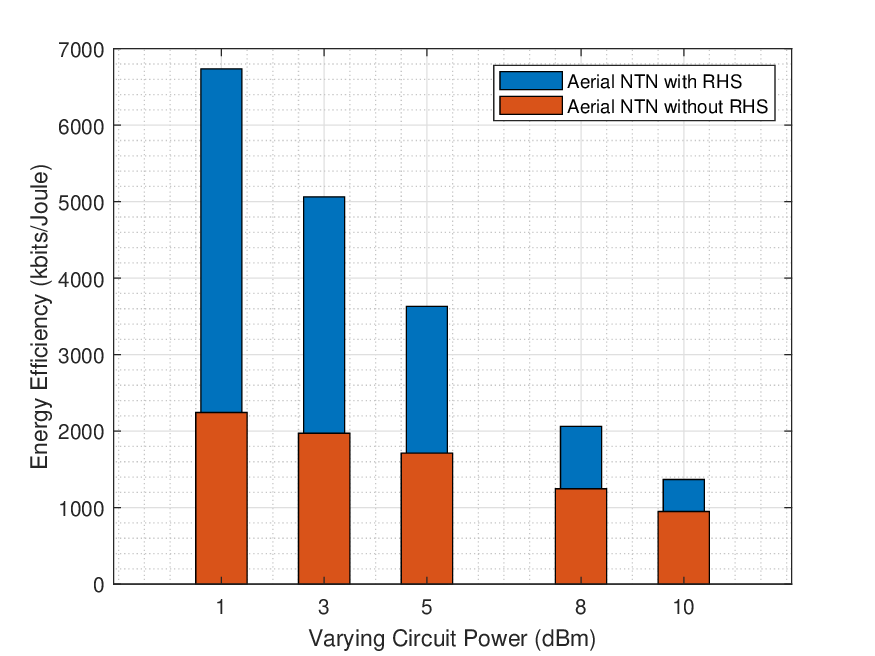}
    \caption{\ac{EE} versus the varying number of circuit power for a fixed transmitted data size.}
    \label{Fig22}
\end{figure}
\section{Design and Analysis}
In this section, a public safety use case analysis is performed to demonstrate the superiority of integrating a \ac{UAV} with \ac{RHS} to improve the \ac{EE} of devices, as shown in Fig. \ref{fig11}. We have considered a disaster region, where users cannot communicate with the outside world, due to damaged cellular infrastructure, and are equipped with a limited power supply. In order to analyze \ac{RHS}, the carrier frequency has been set at 26.2 GHz, having a transmission bandwidth of 120 MHz, where the noise power is set at -174 dBm and the transmit power is set to 23 dBm. The \ac{RHS} array consists of 4 panels, each with \( 48 \times 8 \) elements, where, the element spacing is \( \Delta x = 2.64 \, \text{mm} \) and \( \Delta y = 5.65 \, \text{mm} \). Each panel contains 8 feeds. The \ac{RHS} panels are arranged spatially with precise offsets and are integrated with a \ac{UAV} placed at a height of \( 1000 \, \text{m} \). A total of $K$ users are located on the ground. Their azimuth angles are uniformly distributed between \(-50^\circ\) and \(50^\circ\), and their elevation angles are set at \( 0^\circ \). For each user, the object wave propagation vector is calculated, and a holographic interference pattern is generated over the \ac{RHS} surface. These individual patterns are superimposed and thresholded (binarized) to produce the final active holographic pattern used for transmission. The beamforming matrix is then formed by combining these patterns with distance-dependent phase delays from each \ac{RHS} element to each feed. Similar, to \cite{jamshed2025non}, an optimization problem has been formulated and solved using iterative methods to maximize the \ac{EE} of the system.

In Fig. \ref{Fig22}, a comparison has been conducted on the efficacy of integrating a \ac{UAV} with \ac{RHS} and a \ac{UAV} without \ac{RHS} in relation to the varying circuit power of the system, while the data transmitted by each user remains constant at 50 kbits, and the number of users is fixed to 3. As illustrated in Fig. \ref{Fig22}, the augmentation in the circuit power correlates with a decrease in the \ac{EE} of the system, culminating in the formation of an exponential curve. When comparing with a reliance solely on \ac{UAV}, the integration of \ac{UAV} with \ac{RHS} shows notable improvement in \ac{EE}, showcasing the superiority of this combination.  By leveraging the \ac{UAV} mobility, the \ac{RHS} can be dynamically positioned to maintain optimal line-of-sight conditions with users, minimizing path loss and reducing the required transmit power. The holographic surface enables precise beam shaping and directional transmission, allowing energy to be focused only where needed, thereby avoiding unnecessary radiation and interference. 


\section{Ideas for Future Consideration}
Looking ahead, several innovations are anticipated to further strengthen the utilization of \ac{RHS} and \ac{NFC} for \ac{NTN}, enhancing its scalability, \ac{EE}, mobility support, and seamless integration into future networks. 

\subsection{Spherical-Wavefront Beamforming} \ac{NFC} over both high-frequency and low-\ac{THz} links requires beamforming that concentrates energy at a specific point in range–azimuth–elevation space, rather than along a single far-field direction. Recent tri-hybrid transmit architectures integrate an electromagnetic hologram, an analog RF precoder, and a digital baseband stage. In this setup, the holographic surface generates a coarse focal region, which the RF and baseband stages then refine to enable sub-millisecond multi-user separation \cite{liu2025tritimescalebeamformingdesigntrihybrid}. Early channel sounding studies with \ac{ELAA} show that near-field radiation deviates significantly from plane-wave assumptions. As a result, new range–angle codebooks and compressed pilot designs are needed to maintain training overhead that scales sub-linearly with the number of antenna elements \cite{astrom2024riscellularnetworks}. Extending these ideas, researchers have demonstrated that a printed grid of loop resonators placed behind a holographic panel can share a common focal point with a 275 GHz plasmonic feed. This suggests the possibility of seamless dual-frequency near-field tunnels for applications such as short-range payments or authentication in future 6G terminals.


\subsection{Multiscale Metasurface Architectures} Delivering \ac{NFC}, both at lower frequencies and \ac{mmWave}/\ac{THz} links from a single aperture requires multiscale unit cells that resonate at both centimeter and micrometer wavelengths. Optical research groups have demonstrated lattices where sub-$\lambda$/20 spiral inductors couple efficiently to high-frequency evanescent fields, while neighboring $\lambda$/2 split-ring patches can steer 100 GHz radiation with minimal mutual interference, establishing the concept of dual-channel holography on a single surface. A complementary microwave prototype combines 3.5 and 28 GHz \ac{RIS} elements on a shared FR-4 substrate, enabling independent beam steering in both frequency bands~\cite{10777622}. However, it also reveals that isolating the bias network and managing thermal gradients become critical challenges when switching speeds exceed several tens of kilohertz. As a result, future designs must integrate electromagnetic synthesis with thermal management and low-leakage switching circuits.


\subsection{NFC-Adaptive ELAA Arrays} Mass-market adoption depends on shrinking large-scale holographic panels into battery-compatible form factors without compromising near-field control. Graphene plasmonics offers a promising route, where centimeter-scale chips patterned with nano-ribbons have demonstrated over 20 dB of gain near 320 GHz, despite element spacings below $\lambda$/100, proving that, in principle, a meter-scale aperture can be folded into a handheld device. In parallel, FPGA-based control can combine flexibility with \ac{EE} to perform real-time beam switching for digital metasurface antenna with microsecond latencies and only a few milliwatts per element \cite{petrou2022asic}. The next research step is developing \ac{NFC}-adaptive control logic: when a device battery level is low, the \ac{ELAA} should fall back to passive \ac{NFC} backscatter; under higher power conditions, it should enable active \ac{THz} focusing or distributed \ac{MIMO} across multiple compact tiles. Realizing such adaptive arrays will require advances in nanosecond-response liquid-metal pixels and sub-microsecond PIN-diode drivers, which remain open research challenges \cite{SADON2024100483}.


\subsection{Mobility-Aware Holographic Relays} Mounting \ac{RHS} on \ac{NTN} transforms them into agile relays capable of steering beams around obstacles or concentrating capacity where user density is highest. Recent optimization studies show that jointly planning \ac{UAV} trajectories, transmit power, and active \ac{RIS} phase shifts can significantly improve spectral efficiency while keeping per-flight energy consumption within feasible limits \cite{tyrovolas2024energy}. However, these systems remain highly sensitive to airframe vibrations and double Doppler effects, especially when both the drone and the user terminals are in motion. To address this, transformer-based neural trackers that analyze successive beam-training sweeps have started to mitigate dual mobility, maintaining beam alignment over distances of tens of meters with millisecond-level update cycles. Key remaining challenges include enabling high capacity Ka-band or free-space optical backhaul, currently limited to sub-gigabit per second rates for most small \ac{UAV}, and developing rapid calibration techniques that correct phase distortions caused by rotor-induced jitter within each frame interval.

\subsection{Near-Field Security Protocols} Focusing energy into tight 3D bubbles provides an intuitive form of physical-layer secrecy, yet adversarial metasurfaces have proven that a foil-thin \ac{RIS} placed meters away can invisibly distort or siphon a beam, creating a metasurface-in-the-middle attack. Extending \ac{NFC} standards into such environments, therefore, requires cryptographic handshakes that include range–angle fingerprints derived from the instantaneous near-field impulse response, binding a session to its unique spatial channel, and blocking rogue reflectors. Meanwhile, \ac{3GPP} Release-19 discussions postponed full \ac{RIS} specification partly because secure, low-latency control channels for millions of passive elements remain undefined, signaling that signed element-configuration messages, mutual attestation, and key-derivation from near-field reciprocity are essential standardization topics for the coming study cycles~\cite{10.1145/3507657.3529660}.

\section{Conclusion} 
This study proposes an integrated approach to use \ac{NFC}, \ac{RHS}, and \ac{NTN} as a base for 6G networks. The combination of \ac{RHS} beamforming precision, \ac{NFC} spatial awareness, and the dynamic reach of \ac{NTN} platforms form the solution geared towards \ac{EE}, ultra-low latency, and ubiquitous coverage. The public safety use case verified real world effectiveness of \ac{RHS}-assisted \ac{UAV}, ensuring network resiliency and performance in constrained environments. Effective secure control of the intelligent surfaces, real-time beam steering, scalable adaptability with minimal interference attenuation were scoped as defining core problem areas. Advanced configuration design logic for adaptive metasurfaces that incorporate mobility-aware relaying need immediate attention to actualize operational scale deployable systems within these constraints. This combination provides initial steps towards building smarter future communication systems that integrate resilient low power high capacity needs able to serve multi-purpose functions in integrated \ac{TN} and \ac{NTN} operations domains.

\bibliographystyle{IEEEtran}

\bibliography{IEEEabrv,BibRef}

\end{document}